\documentclass[aps,showpacs,amsmath,amssymb,twocolumn,nofootinbib]{revtex4-1}

\usepackage{subfigure}
\usepackage{graphicx}
\usepackage{dcolumn}
\usepackage{bm}
\usepackage{color}
\usepackage[colorlinks=true,pdfstartview=FitV,linkcolor=blue,citecolor=blue,urlcolor=blue]{hyperref}
\usepackage[mathlines]{lineno}
\usepackage[dvipsnames]{xcolor}
\usepackage{amsmath}

\everymath{\displaystyle}
\begin{document}

\newcommand{\be}{\begin{equation}}
\newcommand{\ee}{\end{equation}}
\newcommand{\bea}{\begin{eqnarray}}
\newcommand{\eea}{\end{eqnarray}}

\title{Dark Photon Dark Matter in the minimal $B-L$ Model}

\author{Gongjun Choi,$^{1}$}
\thanks{{\color{blue}gongjun.choi@gmail.com}}

\author{Tsutomu T. Yanagida,$^{1,2}$}
\thanks{{\color{blue}tsutomu.tyanagida@ipmu.jp}}

\author{Norimi Yokozaki,$^{3}$}
\thanks{{\color{blue}n.yokozaki@gmail.com}}

\affiliation{$^{1}$ Tsung-Dao Lee 
Institute, Shanghai Jiao Tong University, Shanghai 200240, China}

\affiliation{$^{2}$ Kavli IPMU (WPI), UTIAS, The University of Tokyo,
5-1-5 Kashiwanoha, Kashiwa, Chiba 277-8583, Japan}

\affiliation{$^{3}$ Theory Center, IPNS, KEK, 1-1 Oho, Tsukuba, Ibaraki 305-0801, Japan}
\date{\today}

\begin{abstract}
The extension of the Standard model (SM) with three heavy right handed neutrinos, a complex scalar and the gauged $U(1)_{\rm B-L}$ symmetry (the minimal $B-L$ model) is considered the most compelling minimal one: the presence and the out-of-equilibrium decay of the heavy right handed neutrinos can account for the small masses of the active neutrinos and the baryon asymmetry of the universe. A natural accompanying question concerns whether the minimal $B-L$ model can naturally accommodate an interesting dark matter (DM) candidate. We study the possibility where the current DM population is explained by the gauge boson of $U(1)_{\rm B-L}$ symmetry. We discuss how the minimal set-up originally aimed at the seesaw mechanism and the leptogenesis is connected to conditions making the gauge boson promoted to a DM candidate.
\end{abstract}

\maketitle
\section{Introduction}  
Among various different extensions of the Standard model (SM), the model incorporated with the gauged $U(1)_{\rm B-L}$ symmetry, three heavy right-handed neutrinos ($\overline{N}_{i=1,2,3}$) and a complex scalar ($\Phi$) is regarded as one of the most compelling minimal one (the minimal $B-L$ model): the model can answer the questions about origins of the tiny active neutrino mass and the observed baryon asymmetry in the Universe. The condensation of the complex scalar $\Phi$ induces the spontaneous breaking of $U(1)_{\rm B-L}$ while simultaneously imposing masses to the three right-handed neutrinos. For a sufficiently large vacuum expectation value (VEV) of $\Phi$, helped by the interaction between the active and right-handed neutrinos via Dirac mass terms, the heaviness of the right handed neutrinos makes mass eigenvalues of the active neutrinos very tiny (seesaw mechanism)~\cite{Yanagida:1979as,GellMann:1980vs,Minkowski:1977sc}. Moreover, the out-of-equilibrium decay of heavy right handed neutrinos can seed the primordial lepton asymmetry which is to be converted into the baryon asymmetry later thanks to the sphaleron transition (leptogenesis)~\cite{Fukugita:1986hr,Buchmuller:2005eh}.

Motivated by the powerful capability of the minimal $B-L$ model to address the aforementioned two problems, several works have been done regarding the question about a dark matter (DM) candidate in the model~\cite{Choi:2020nan,Kusenko:2010ik,Khalil:2008kp,Ibe:2016yfo,Kanemura:2014rpa,Singirala:2017cch,Abdallah:2019svm,Lindner:2011it,Okada:2016gsh,Biswas:2016ewm,Wang:2015saa,Basak:2013cga,Bandyopadhyay:2017bgh,DeRomeri:2017oxa,Okada:2010wd,Kanemura:2011vm,Seto:2016pks,Bandyopadhyay:2018qcv,Choi:2020kch,Han:2020oet,Okada:2020evk,Lindner:2020kko,FileviezPerez:2019cyn,Das:2019pua}. In most of the cases, either fermion or scalar in the model is taken to be a DM candidate. However, from the theoretical point of view, there is no any compelling reason to allow for a huge separation of mass scales between each $\overline{N}_{i}$ as well as to consider a light or intermediate scale scalar mass. In addition, introducing extra degrees of freedom other than the complex scalar and three right-handed neutrinos into the model to explain DM spoils the minimality of the minimal $B-L$ model. Along this line of reasoning, one question that naturally arises concerns the possibility where the gauge boson ($A_{\mu}^{'}$) of $U(1)_{\rm B-L}$ plays the role of DM candidate in the model.

There exist several interesting observations that can logically support the choice of $A_{\mu}^{'}$ as the DM candidate. First of all, no additional set-up is needed to be consistent with the massive DM. The breaking of $U(1)_{\rm B-L}$ is unavoidable for the seesaw mechanism and the leptogenesis to operate and thus the non-vanishing mass of $A_{\mu}^{'}$ is the consequence of the small active neutrino masses and the baryon asymmetry in the model. Secondly, the explanation for the weak non-gravitational interaction becomes very economical. The null observation of DM to date can serve as the compelling evidence for a very weak strength of DM's non-gravitational interaction with ordinary matters in the low energy scale if there is. Now that the non-gravitational interaction with which $A_{\mu}^{'}$ is involved is uniquely described by the gauge interaction, it suffices to assume a small enough gauge coupling ($g_{\rm B-L}$) of $U(1)_{\rm B-L}$ for accomplishing the required weak non-gravitational interaction without further suppression of other coupling constants. Thirdly, thanks to the relation $m_{A'}=2g_{\rm B-L}V_{\rm B-L}$, a small enough $g_{\rm B-L}$ so taken can prevent too heavy DM mass, readily accepting a sufficiently large vacuum expectation value (VEV) of the complex scalar $\langle\Phi\rangle\equiv V_{\rm B-L}/\sqrt{2}$ by which the seesaw mechanism and the leptogenesis should be necessarily accompanied for their success. 

In accordance with the above insight, in this paper, we study scenarios where $A_{\mu}^{'}$ of $U(1)_{\rm B-L}$ is responsible for the DM population today. As we shall see, the mass regime of $A_{\mu}^{'}$ as a DM candidate is keV-scale to be consistent with the existing experimental data. Exceeding the upper bound $\sim100{\rm eV}$ of the mass of DM with an identical temperature to the SM thermal bath, keV-scale necessitates the assumption for a cooler temperature of $A_
{\mu}^{'}$ than that of the SM thermal bath. Therefore, in our model, we intentionally suppress the operators which can potentially make $A_{\mu}^{'}$ be thermalized by the SM thermal bath. For this purpose, we assume a negligibly small (i) kinetic mixing between $A_{\mu}^{'}$ and the hypercharge gauge boson in the SM, and (ii) mixing between $\Phi$ and the SM Higgs SU(2) doublet even if presence of those operators are allowed by the symmetry of the model. We invoke the interaction between $\Phi$ and $\bar{N}_{i}$ responsible for $\bar{N}_{i}$'s mass, i.e. $\Phi\bar{N}_{i}\bar{N}_{i}$ for creating the dark sector system isolated from the SM sector. As we shall see, there can be several intriguing possibilities where the conspiracy between the reheating temperature, the right handed neutrino mass, and the scalar potential in the model enables $A_{\mu}^{'}$ to be identified with the mysterious DM successfully . Thereby, in this paper, the minimal $B-L$ model will be shown to be able to answer the three key questions (DM, the active neutrino mass, baryon asymmetry) that the SM cannot address alone.

\section{The minimal $B-L$ Model}
\label{sec:model}
As the minimal set-up for implementing the seesaw mechanism and the leptogenesis, we extend the gauge sector of the SM by adding the new gauge symmetry $U(1)_{\rm B-L}$ for which the conserved charge ($Q_{\rm B-L}$) is the difference between a baryon number ($B$) and a lepton number ($L$) of a particle charged under $U(1)_{\rm B-L}$. In terms of particle contents, on top of the SM particles we consider a complex scalar $\Phi$(-2) and three right-handed neutrinos $\overline{N}_{i=1,2,3}$(+1) with the corresponding $Q_{\rm B-L}$ specified in each parenthesis. We further impose $Q_{\rm B-L}\!\!=\!\!+1/3(-1/3)$ to the left handed Weyl fields for the quarks (anti-quarks) and $Q_{\rm B-L}\!\!=\!\!-1(+1)$ to the left handed Weyl fields for the leptons (anti-leptons) so that gauge anomaly cancellations for $U(1)_{\rm B-L}\times G_{\rm SM}^{2}$, $U(1)_{\rm B-L}^{3}$ and $U(1)_{\rm B-L}\times[\rm gravity]^{2}$ are realized where $G_{\rm SM}$ is an element of the SM gauge group.

The interactions with which the complex scalar $\Phi$ is involved in the model are given by the following operators
\be
\mathcal{L}_{\Phi}=m_{\Phi}^{2}|\Phi|^{2}-\lambda|\Phi|^{4}-\left(\sum_{i=1}^{3}\frac{1}{2}y_{i}\Phi\overline{N}_i \overline{N}_i +{\rm h.c.}\right)\,,
\label{eq:LPhi}
\ee
where the terms in the parenthesis is written in the basis of $\overline{N}_{i}$ where the Yukawa coupling matrix is a diagonal one. The acquisition of VEV of $\Phi$ minimizing the first two terms ($-V(\Phi)$) induces the spontaneous breaking of $U(1)_{\rm B-L}$ and imposes mass $m_{\bar{N},i}\simeq y_{i}V_{\rm B-L}/\sqrt{2}$ to the right handed neutrinos. On breaking of $U(1)_{\rm B-L}$, the gauge boson $A_{\mu}^{'}$ becomes massive with the mass $m_{A'}=2g_{\rm B-L}V_{\rm B-L}$ by absorbing the Nambu-Goldstone mode $\theta$ of $\Phi\equiv(\phi/\sqrt{2}+V_{\rm B-L}/\sqrt{2})e^{i\theta/V_{\rm B-L}}$. The $U(1)_{\rm B-L}$ breaking scale can be written as $V_{\rm B-L}\simeq m_{\Phi}/\sqrt{\lambda}= m_{\phi}/\sqrt{2\lambda}$ where $m_{\phi}$ is the mass of $\phi$.

As for the relation between $U(1)_{\rm B-L}$ breaking scale ($V_{\rm B-L}$) and a reheating temperature ($T_{\rm RH}$), we notice that for $V_{\rm B-L}<T_{\rm RH}$, the $U(1)_{\rm B-L}$ symmetry restoration is likely to happen at the reheating era even if the breaking of $U(1)_{\rm B-L}$ took place before the reheating era. To avoid this complicated situation and make our analysis simpler, we assume the following relation from now on,
\be
{\rm (1a)}:\quad V_{\rm B-L}>T_{\rm RH}\quad\rightarrow\quad\frac{m_{\phi}}{\sqrt{2\lambda}}>T_{\rm RH}\,.
\label{eq:VT}
\ee
In addition, to make $U(1)_{\rm B-L}$ be in the broken phase at the reheating era, we impose the following condition 
\be
{\rm (2a)}:\quad m_{\Phi}>\mathcal{H}(a=a_{\rm RH})\quad\rightarrow\quad\sqrt{m_{\Phi}M_{P}}>T_{\rm RH}\,,
\label{eq:brokenU(1)}
\ee
where $\mathcal{H}\equiv\dot{a}/a$ is the Hubble expansion rate. With Eq.~(\ref{eq:VT}) and Eq.~(\ref{eq:brokenU(1)}), the right-handed neutrinos are already massive at the reheating era. For Sec.~\ref{sec3} and Sec.~\ref{sec4}, we make it sure that conditions (1a) and (2a) in Eq.~(\ref{eq:VT}) and Eq.~(\ref{eq:brokenU(1)}) are satisfied for the consistent parameter spaces.

In this work we shall assume $V_{\rm B-L}>\!\!>V_{\rm EW}$, and so $\overline{N}_{i}$s and $\phi$ become irrelevant to the low energy physics unless certain special arrangements to make their comoving number densities conserved. Thus we attend to the last non-SM particle in the model, $A_{\mu}^{'}$, in searching for a DM candidate in the minimal $B-L$ model. Taking $A_{\mu}^{'}$ as the DM candidate, we see the necessity of the suppression for the decay of $A_{\mu}^{'}$ to SM fermion-antifermion pairs. For this, we demand $\Gamma(A_{\mu}^{'}\rightarrow f+\bar{f})\lesssim(13.8{\rm Gyr})^{-1}$ to obtain $m_{A'}^{3}<10^{-40}V_{\rm B-L}^{2}{\rm GeV}$ where $f$ is a SM fermion with its mass satisfying $m_{A'}\gtrsim2m_{f}$ and $13.8{\rm Gyr}$ is the age of the universe. Furthermore, for $m_{A'}>1{\rm MeV}$, applying the lower bound of the lifetime $\sim10^{24}{\rm sec}$ for the decaying DM obtained based on the diffuse photon spectra data~\cite{Essig:2013goa}, we obtain $m_{A'}^{3}<10^{-46}V_{\rm B-L}^{2}{\rm GeV}$. We find that $m_{A'}\lesssim1{\rm MeV}$ can readily satisfy these constraints for $V_{\rm B-L}$ of our interest ($V_{\rm B-L}\lesssim M_{P}$) with $M_{P}=2.4\times10^{18}{\rm GeV}$ the reduced Planck mass.\footnote{In addition, the choice of the mass regime $m_{A'}\lesssim1{\rm MeV}$ makes the decay $A_{\mu}^{'}\rightarrow e^{-}+e^{+}$ kinematically suppressed so as to preclude the production of the resultant $\gamma$-ray flux as the final state radiation. Note that the the possible decay $A_{\mu}^{'}\rightarrow\gamma\gamma\gamma$ is highly suppressed as well due to the assumption of the negligibly small kinetic mixing between $A_{\mu}^{'}$ and $\gamma$ and the small $g_{\rm B-L}$.} Therefore, we restrict ourselves to the mass regime $m_{A'}\lesssim1{\rm MeV}$ as long as $m_{A'}$ is consistent with the lower bound on the warm DM mass, i.e. $\mathcal{O}(1-10){\rm keV}$ coming from the Lyman-$\alpha$ forest observation~\cite{Irsic:2017ixq}. With this mass regime in mind, we find that the life time constraint above $m_{A'}^{3}<10^{-40}V_{\rm B-L}^{2}{\rm GeV}$ gives us $g_{\rm B-L}<5\times10^{-18}\times(m_{A'}/1{\rm keV})^{-1/2}$, which is sufficiently small to invalidate gauge interaction-induced thermalization between the SM and the dark sector.

\begin{figure*}[htp]
  \centering
  \hspace*{-5mm}
  \subfigure{\includegraphics[scale=0.435]{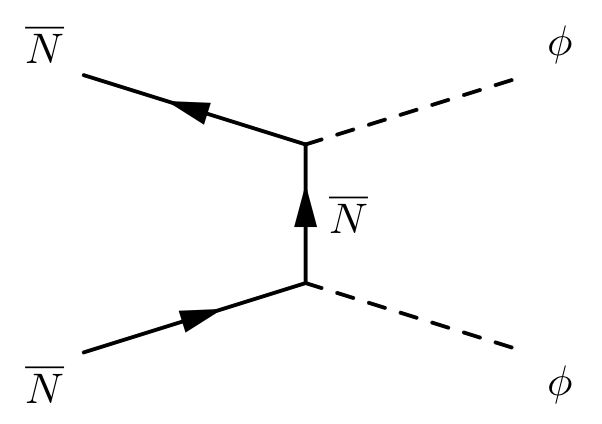}}\quad\quad
  \subfigure{\includegraphics[scale=0.435]{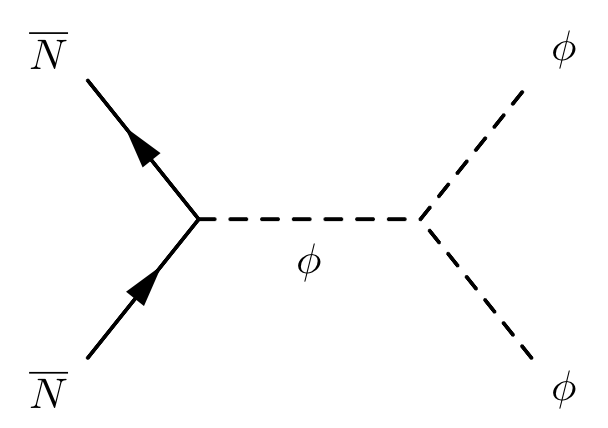}}\quad\quad
   \subfigure{\includegraphics[scale=0.435]{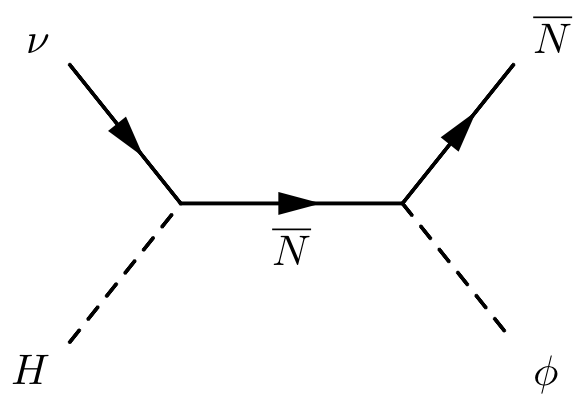}}
  \caption{The diagrams for $\phi$-production processes. $\overline{N}$, $\phi$, $H$ and $\nu$ denote the right-handed neutrino responsible for $\phi$-production, the radial component of $\Phi$, the SM SU(2) doublet Higgs and the active neutrino in the SM, respectively. In this paper, we consider the case where the first $\overline{N}$-$\overline{N}$ $t$-channel scattering is the dominant one in $\phi$-production.}
  \vspace*{-1.5mm}
\label{fig}
\end{figure*}

With the mass scale of the DM candidate specified above, we realize that $A_{\mu}^{'}$ should never reside in the SM thermal bath unless the model can accommodate an exotic late time entropy production after DM candidate gets out of the SM thermal bath. In the absence of such a late time entropy production, it is demanded that $A_{\mu}^{'}$ be produced from something else cooler than the SM thermal bath to avoid to overclose the universe. Then how could we establish such a dark sector system? As we already mentioned in the introduction, in our model the negligibly small kinetic mixing of $U(1)$ gauge bosons and the mixing between scalars are assumed for preventing $A_{\mu}^{'}$ from being in equilibrium with the SM thermal bath. In addition, the aforementioned tiny strength of $g_{\rm B-L}$ makes it very difficult to have the gauge interaction-induced $A_{\mu}^{'}$ production for the time before the matter-radiation equality. For putting $A_{\mu}^{'}$ in a system cooler than the SM sector, we consider the case where the visible sector of the model is composed of the SM particles and the three right-handed neutrinos while the dark sector of the model includes $\Phi$ and $A_{\mu}^{'}$. $\Phi$ shall serve as the assistant to have $A_{\mu}^{'}$ in the system cooler than the SM sector. We attend to terms in the parenthesis in Eq.~(\ref{eq:LPhi}) as the last portal to the dark sector from the SM sector. Namely, we consider the situation where $\phi$ is produced non-thermally based on a process due to the operator $\sim\Phi\overline{N}_{i}\overline{N}_{i}$ and initiates the desired isolated dark sector system.

Depending on whether the right-handed neutrinos responsible for the leptogenesis is heavier than $T_{\rm RH}$ or not, there can be two different ways of producing the primordial lepton asymmetry: the thermal leptogenesis and the non-thermal leptogenesis. For the former, at least one of right-handed neutrinos for the leptogenesis is lighter than $T_{\rm RH}$ and $T_{\rm RH}\gtrsim10^{9}{\rm GeV}$ is required. For the latter, all the right-handed neutrinos for the leptogenesis are heavier than $T_{\rm RH}$ and  $T_{\rm RH}\gtrsim10^{6}{\rm GeV}$ is required~\cite{Buchmuller:2005eh}. We notice that the it suffices to rely on two right handed neutrinos for successful working of both the seesaw mechanism and the leptogenesis~\cite{Frampton:2002qc}. We invoke this point to make our analysis simpler so that we attribute the seesaw mechanism and the leptogenesis to two right-handed neutrinos $\overline{N}_{1}$ and $\overline{N}_{2}$\footnote{Accordingly, one mass eigenvalue of the three light neutrinos is also suppressed, which is, however, consistent with all neutrino oscillation experiments.} while assuming the third one $\overline{N}\equiv\overline{N}_{3}$ irrelevant to the two mechanisms by suppressing the operator $\mathcal{O}= Y_{\nu}LH^{\dagger}\overline{N}$.\footnote{For the suppression, however, we still keep the coupling constant $Y_{\nu}$ non-vanishing so that $\overline{N}$ can still be in thermal equilibrium with the SM thermal bath when $\phi$-production is most active, i,e. when $T_{\rm SM}\simeq m_{\bar{N}}$.} Here $L$ is the SM $SU(2)$ lepton doublet. Armed with this set-up, for simplicity of our upcoming analysis, in this paper we consider the case in which the following operator is totally responsible for initiating the isolated dark sector system
\be
\mathcal{O}=\frac{y}{2\sqrt{2}}\phi\overline{N}\overline{N}\,,
\label{eq:ODS}
\ee
where $y\equiv y_{3}$ is defined and $y_{3}$ is from Eq.~(\ref{eq:LPhi}). To this end, we restrict ourselves to the following two cases: 
\begin{itemize}
    \item $m_{\bar{N},1}, >T_{\rm RH}>m_{\bar{N}}>m_{\bar{N},2}\gtrsim10^{9}{\rm GeV}$ (thermal leptogenesis)
    \item $m_{\bar{N},1},m_{\bar{N},2}>T_{\rm RH}>m_{\bar{N}}$ (non-thermal leptogenesis)
\end{itemize}
Here $m_{\bar{N},i}$ ($m_{\bar{N}}$) is the mass of $\overline{N}_{i}$ ($\overline{N}$). For the first case, $\overline{N}_{2}$ is in the SM thermal bath due to the interaction $\sim LH^{\dagger}\overline{N}_{2}$ and its out-of-equilibrium decay seeds the primordial lepton asymmetry. Because of $m_{\bar{N}}>m_{\bar{N},2}$, $\phi$-production due to $\overline{N}$ dominates over $\overline{N}_{2}$. For the second case, both $\overline{N}_{1}$ and $\overline{N}_{2}$ are not in equilibrium with the SM thermal bath, but still responsible for the generation of the primordial lepton asymmetry via the out-of-equilibrium decay. For the temperature below $T_{\rm RH}$, they become decoupled in the theory and so irrelevant to $\phi$-production. Thus in either case, we only need to care about $\overline{N}$ concerning $\phi$-production process. Aside from the above relation between mass scales, we further assume
\be
m_{\bar{N}}>m_{\phi}\,,
\ee
in order that the decay of $\phi$ to a pair of $\overline{N}$ is kinematically suppressed so that we need not worry about disappearance of $\phi$ before the isolated dark sector system containing $A_{\mu}^{'}$ is generated.

In Fig.~\ref{fig}, we show the possible tree-level $\phi$-production processes thanks to the operator in Eq.~(\ref{eq:ODS}). Since the operator $\sim LH^{\dagger}\overline{N}$ is assumed suppressed in the model, neglecting the third process is justified. In addition, as mentioned already, we restrict our interest to the case where the operator in Eq.~(\ref{eq:ODS}) is the unique portal to the dark sector for simplicity.\footnote{In principle, all the three diagrams in Fig.~\ref{fig} should be summed for a generic discussion and analysis for $\phi$-production which is beyond the scope of this paper.} For this, we demand $\Gamma_{t-{\rm ch}}>>\Gamma_{s-{\rm ch}}$ where $\Gamma_{t-{\rm ch}}$ ($\Gamma_{s-{\rm ch}}$) is the interaction rate corresponding to the first (second) diagram in Fig.~\ref{fig}. This condition is converted into
\be
\left(\frac{y}{2\sqrt{2}}\right)^{4}T_{\rm SM}\,\,>\!\!>\,\,\left(\frac{y}{2\sqrt{2}}\right)^{2}\frac{\lambda^{2}V_{\rm B-L}^{2}}{T_{\rm SM}}\,,
\label{eq:3a}
\ee
where $T_{\rm SM}$ is a SM thermal bath temperature, and $y$ and $\lambda$ are from Eq.~(\ref{eq:ODS}) and Eq.~(\ref{eq:LPhi}). For $T_{\rm SM}<m_{\bar{N}}$, the two processes are kinematically suppressed and so Eq.~(\ref{eq:3a}) remains true as far as it is satisfied for $T_{\rm SM}=m_{\bar{N}}$. Substituting $T_{\rm SM}=m_{\bar{N}}$ into Eq.~(\ref{eq:3a}) yields
\be
{\rm (3a)}:\quad\left(\frac{ym_{\bar{N}}}{2m_{\phi}}\right)^{2}>\!\!>\lambda\quad\rightarrow\quad\left(\frac{y^{2}}{4}\right)>\!\!>\lambda\,.
\label{eq:3a2}
\ee
In Sec.~\ref{sec3} and \ref{sec4}, we make it sure that the condition (3a) given in Eq.~(\ref{eq:3a2}) is always satisfied so that $\phi$-production is dominantly accomplished via the $t$-channel $\overline{N}$-$\overline{N}$ scattering.

After $\phi$-production is completed, the evolution of the dark sector system can be classified by two different cases depending on whether $\phi$ forms a dark thermal bath or not. For the case with the dark thermal bath, as we shall see in Sec.~\ref{sec3}, $\phi$ produces the longitudinal mode of $A_{\mu}^{'}$ (the Nambu-Goldstone mode $\theta$) via its decay and makes it join the dark thermal bath. Thus, $A_{\mu}^{'}$ follows the thermal distribution from the beginning. In contrast, it is possible for the dark sector to lack any thermal bath, being described by the free-streaming $\phi$ and $A_{\mu}^{'}$. In Sec.~\ref{sec4}, we shall study this case without the dark thermal bath and see that $A_{\mu}^{'}$ would be subject to momentum space distribution with much narrower width comparing to the thermal distribution. 

In sum, the minimal $B-L$ model we consider is featured by a $m_{\phi}\lesssim m_{\bar{N}}\lesssim T_{\rm RH}\lesssim V_{\rm B-L}\lesssim M_{P}$, which results in a small gauge coupling constant $g_{\rm B-L}\lesssim10^{-9}$ for the mass regime of $A_{\mu}^{'}$ of our interest. The seesaw mechanism and the leptogenesis are attributed to $\overline{N}_{1}$ and $\overline{N}_{2}$ which are irrelevant to $\phi$-production. The remaining right-handed neutrino $\overline{N}$ is fully responsible for $\phi$-production throughout the $t$-channel scattering thereof. On production of $\phi$, the dark sector could evolve either with or without the dark thermal bath. Stemming from the decay of or the scattering among $\phi$s, $A_{\mu}^{'}$ remains in the low energy with the SM particles as the dark photon dark matter (DPDM) candidate after the heavy degrees of freedoms ($\phi$, $\overline{N}_{i}$ and $\overline{N}$) are integrated-out. 

\section{With Dark Thermal Bath}
\label{sec3}
In this section, we study the case where the early time dark sector is described by the presence of the dark thermal bath differentiated and isolated from the SM thermal bath. The situation we envision in this section is what follows. $\phi$ non-thermally produced from $\overline{N}$-$\overline{N}$ $t$-channel scattering forms the dark thermal bath consisting of $\phi$ and $A_{\mu}^{'}$ (longitudinal component). Then $\phi$ is integrated-out when a dark thermal bath temperature ($T_{\rm DS}$) is comparable to $m_{\phi}$. Since $A_{\mu}^{'}$ itself cannot form the dark thermal bath with its very weak self-interaction, $A_{\mu}^{'}$ starts free-streaming on $\phi$'s disappearance from the dark thermal bath. 

As we shall see in Eq.~(\ref{eq:Yphi}), the non-thermal $\phi$-production rate is inversely proportional to $T_{\rm SM}$. This implies that $\phi$-production is most active at $T_{\rm SM}\simeq m_{\bar{N}}$. For $T_{\rm SM}\lesssim m_{\bar{N}}$, $\overline{N}$ disappears through its decay and pair annihilation, dumping its entropy into the SM thermal bath. 

Once $\phi$ is produced, the dark thermal bath can form when the interaction rate of the decay and inverse decay ($\phi\leftrightarrow A_{\mu}^{'}+A_{\mu}^{'}$) of the relativistic $\phi$ starts to exceed the Hubble expansion rate, i.e. $(m_{\phi}/T_{\rm DS})\times\Gamma(\phi\rightarrow A^{'}\!+\!A^{'})\gtrsim\mathcal{H}$
\be
\frac{m_{\phi}}{T_{\rm DS}}\times(-2)^{4}\frac{g_{\rm B-L}^{4}V_{\rm B-L}^{2}m_{\phi}^{3}}{32\pi m_{A'}^{4}}\gtrsim\frac{T^{2}_{\rm SM}}{M_{P}}\,,
\label{eq:Ajoin}
\ee
where the prefactor $(-2)^{4}$ is due to $B-L$ charge of $\phi$.\footnote{One may wonder other ways to establish the dark thermal bath other than $\phi\leftrightarrow A_{\mu}^{'}+A_{\mu}^{'}$ process. In effect, the processes $\phi+\phi\leftrightarrow\phi+\phi$ and $\phi+\phi\leftrightarrow\theta+\theta$ induced by the quartic term in Eq.~(\ref{eq:LPhi}) and the kinetic term of $\Phi$ can also contribute to the formation of the dark thermal bath. Since the area of the parameter space ($\lambda,m_{\phi}$) causing the dark thermal bath in these ways is confirmed to be smaller than that corresponding to $\phi\leftrightarrow A_{\mu}^{'}+A_{\mu}^{'}$ process, we focus on $\phi\leftrightarrow A_{\mu}^{'}+A_{\mu}^{'}$ process for our discussion of the dark thermal bath formation.} Define $a_{\star}$ to be the scale factor at which $(m_{\phi}/T_{\rm DS}(a_{\star}))\times\Gamma(\phi\rightarrow A^{'}\!+\!A^{'})\simeq \mathcal{H}(a_{\star})$ holds. Now if $T_{\rm DS}(a_{\star})>m_{\phi}$, because the interaction rate is greater than $\mathcal{H}$ for $T_{\rm DS}<T_{\rm DS}(a_{\star})$, the dark thermal bath made up of $\phi$ and $A_{\mu}^{'}$ can form before $\phi$ is integrated-out. Hence, we demand the following as the condition for the formation of the dark thermal bath, i.e. (condition 1b)
\be
({\rm 1b}):\quad T_{\rm DS}(a_{\star})\simeq\left(\frac{\xi^{2}}{16\pi}\lambda m_{\phi}^{2}M_{P}\right)^{1/3}>m_{\phi}\,,
\label{eq:1b}
\ee
where the temperature ratio $\xi\equiv T_{\rm DS}/T_{\rm SM}$ is defined and used.

\begin{figure*}[htp]
  \centering
  \hspace*{-5mm}
  \subfigure{\includegraphics[scale=0.395]{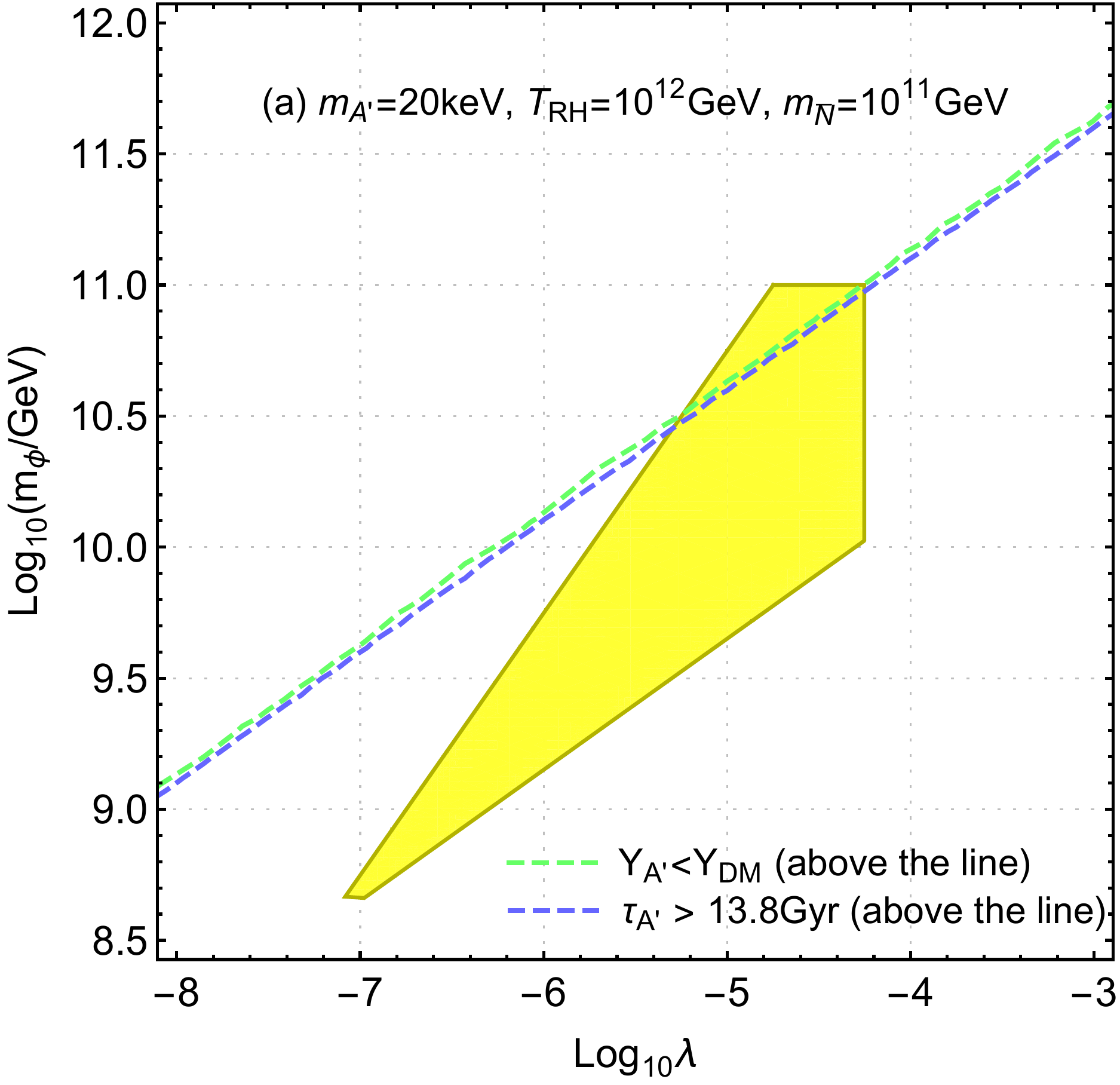}}\qquad
  \subfigure{\includegraphics[scale=0.395]{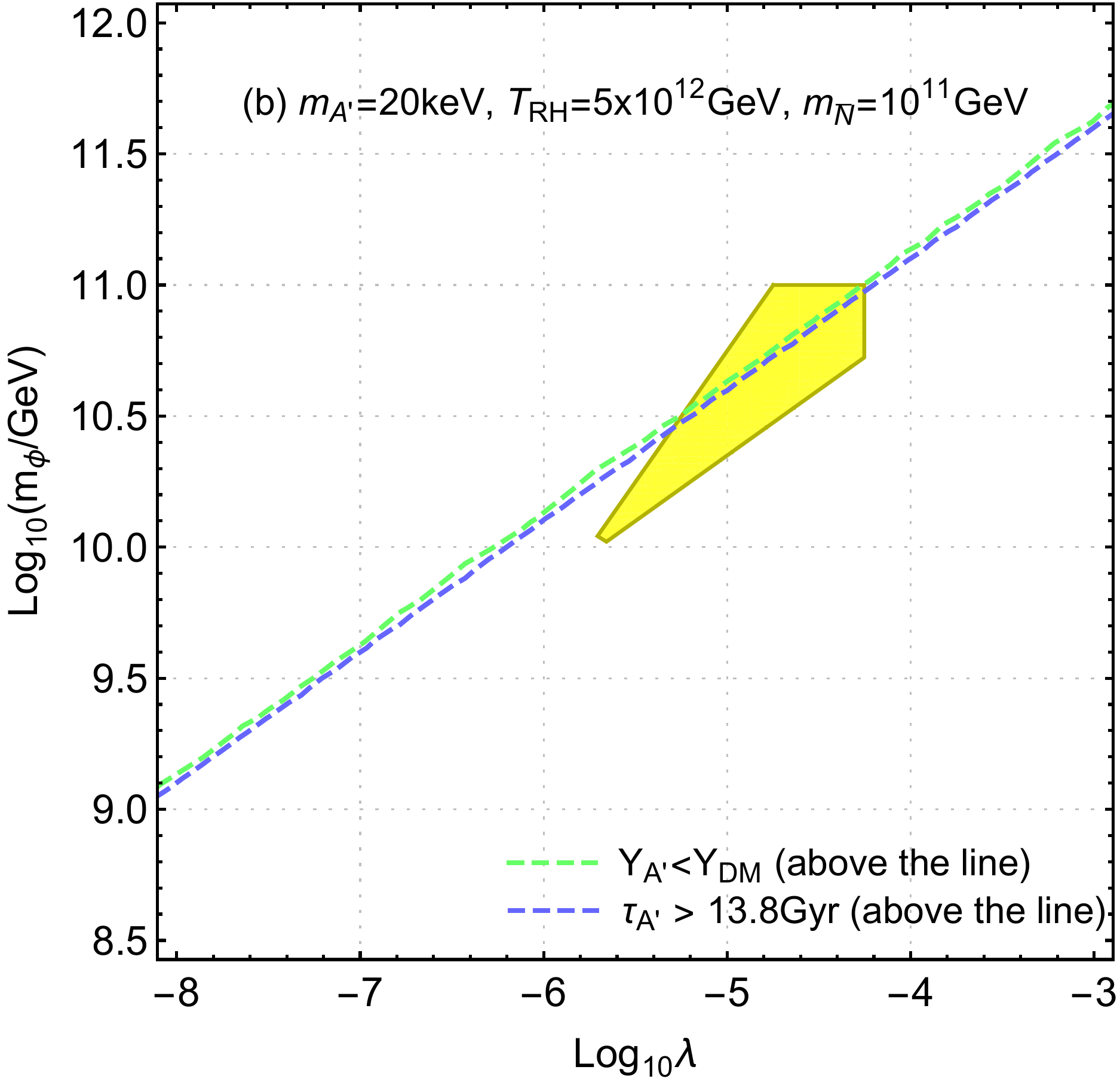}}
  \subfigure{\includegraphics[scale=0.395]{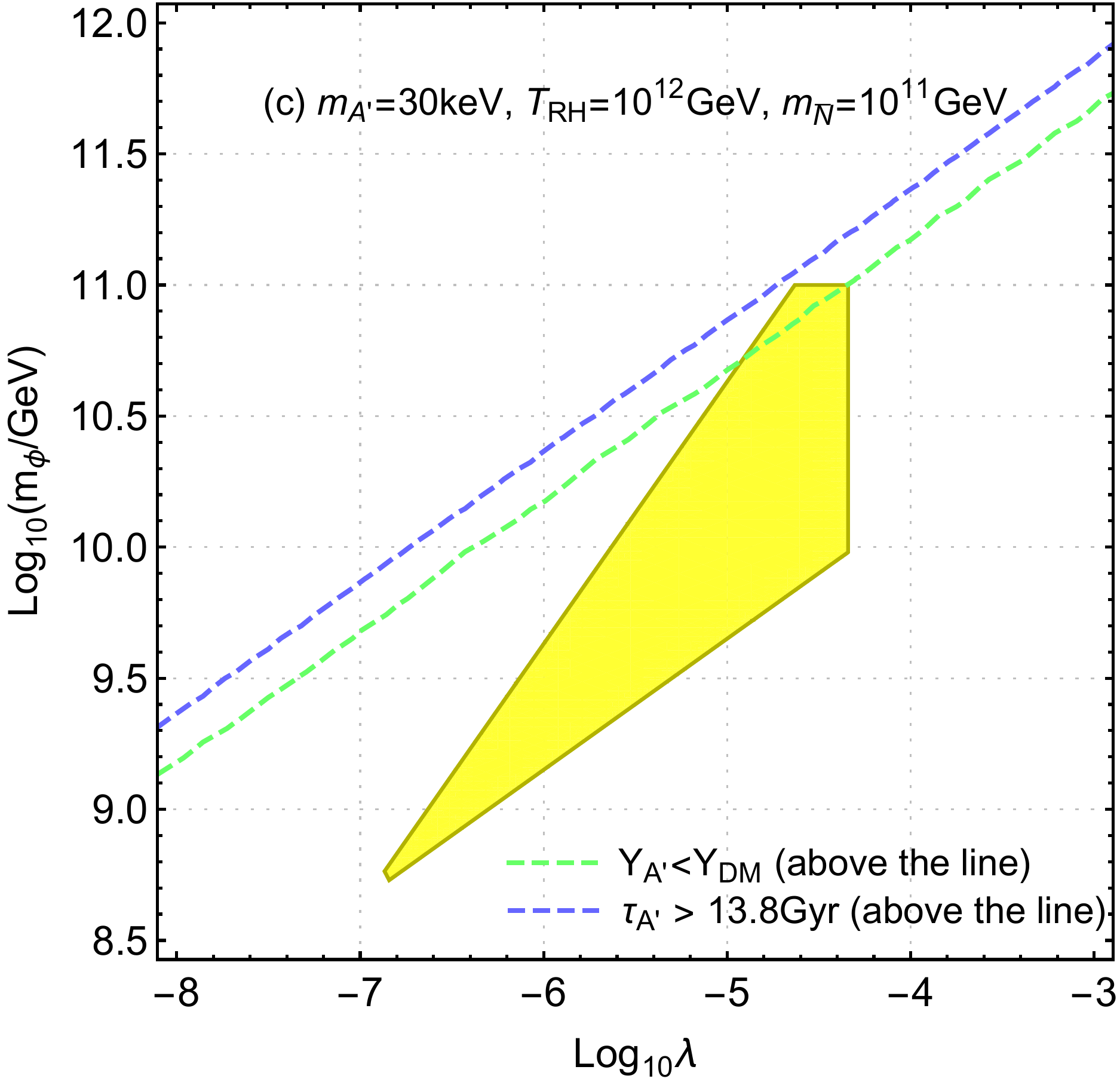}}\qquad
  \subfigure{\includegraphics[scale=0.395]{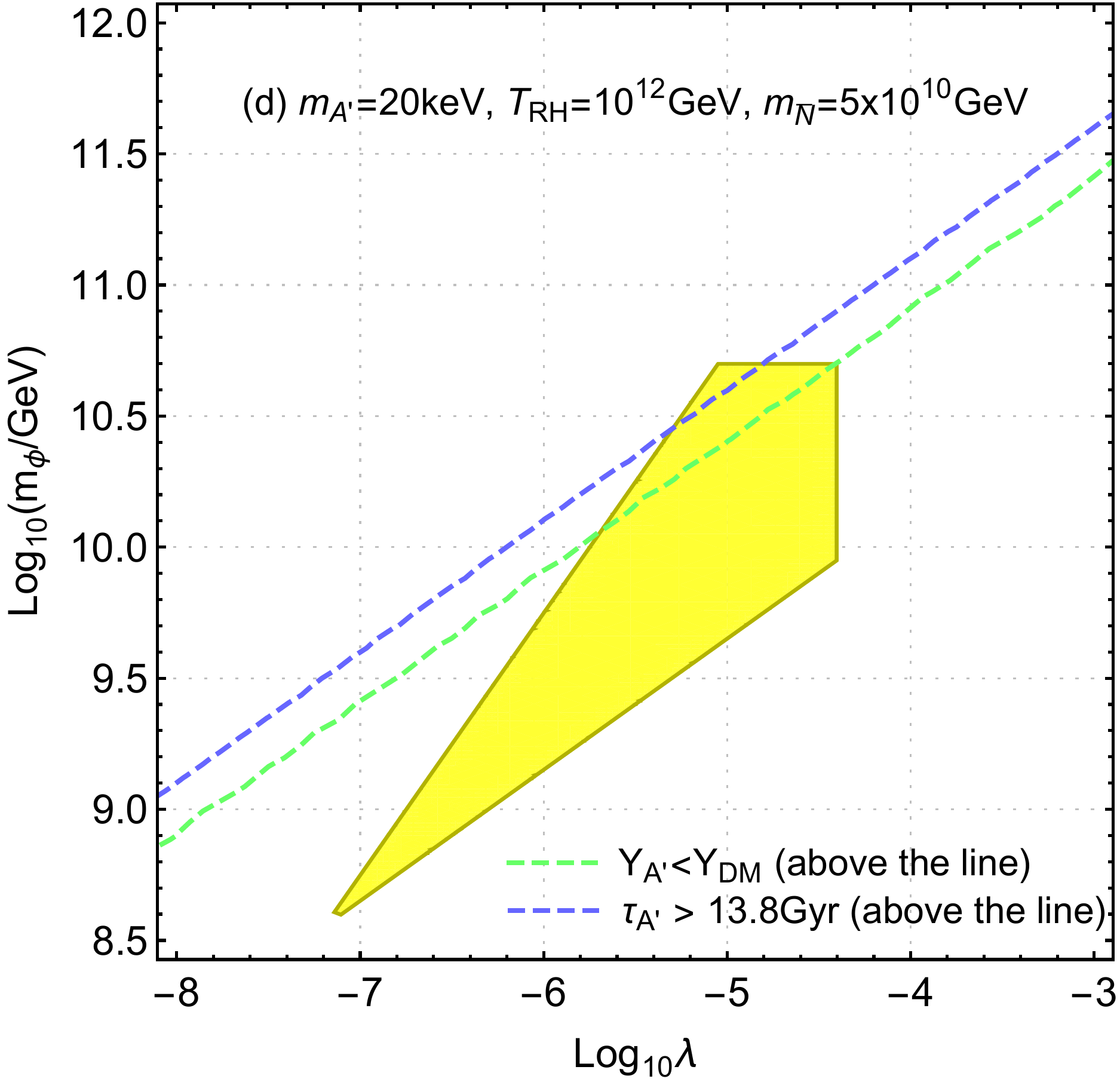}}
  \caption{The constraints on the parameter space of $(\lambda,m_{\phi})$ which is consistent with the conditions (1a), (2a), (3a') and (1b) presented in Eq.~(\ref{eq:VT}), Eq.~(\ref{eq:brokenU(1)}), Eq.~(\ref{eq:3asec3}), Eq.~(\ref{eq:1b}) (yellow shaded region) and  $\tau_{A'}\gtrsim13.8{\rm Gyr}$ (the region above the blue dashed line). Along the green dashed line, $A_{\mu}^{'}$ alone can explain the current DM abundance. The values shown in (a) being fiducial, each plot of (b), (c) and (d) results from variation of one of ($m_{A'},T_{\rm RH}, m_{\bar{N}}$).}
  \vspace*{-1.5mm}
\label{fig:2}
\end{figure*}

As long as $\lambda$ can satisfy both Eq.~(\ref{eq:3a2}) and Eq.~(\ref{eq:1b}) simultaneously for a given set of $(\xi,m_{\phi},m_{\bar{N}})$, the dark thermal bath made up of $\phi$ and $A_{\mu}^{'}$ forms as the system independent of the SM thermal bath. When $T_{\rm DS}\simeq m_{\phi}$ is reached, $\phi$ decays to produce $A_{\mu}^{'}$s and then the free-streaming of $A_{\mu}^{'}$ gets started. Since DPDM considered in this subsection is of the thermal kind, the considered mass regime is  $m_{A'}\in(20{\rm keV},1{\rm MeV})$ to be consistent with Lyman-$\alpha$ forest observation~\cite{Irsic:2017ixq}.\footnote{Based on Appendix.~\ref{appD}, we see that $m_{A'}=20{\rm keV}$ is greater than the lower bound of the mass of $A_{\mu}^{'}$ mapped from $m_{\rm wdm}=5.3{\rm keV}$.} 

Before we constrain the model based on the aforesaid conditions, we discuss the size of $\xi$ and the required strength of $y$ for $A_{\mu}^{'}$ to explain the current DM density. As both $\phi$ and $A_{\mu}^{'}$ are in the thermal bath until $A_{\mu}^{'}$ becomes decoupled, the number density ratio between the two is equal to the ratio of degrees of freedom, i.e. $r\equiv n_{A'}/n_{\phi}=1$.\footnote{Note that only the longitudinal component of $A_{\mu}^{'}$ is produced from $\phi$-decay and becomes DM today.} In order that $A_{\mu}^{'}$ can account for DM population today as the particle that belonged to the dark thermal bath together with $\phi$s, the comoving number density $Y_{\rm DM}$ of DM given in Eq.~(\ref{eq:YDM}) should satisfy~\cite{Khalil:2008kp,Kusenko:2010ik}
\bea
Y_{\rm DM}&\equiv&\frac{n_{A'}}{s_{\rm SM}}=\frac{n_{\phi}}{s_{\rm SM}}\sim\left.\frac{n_{\bar{N}}\Gamma(2\overline{N}\rightarrow2\phi)/\mathcal{H}}{s_{\rm SM}}\right\vert_{T_{\rm SM}\simeq m_{\bar{N}}}\cr\cr
&\simeq&1.5\times10^{5}\times y^{4}\times\left(\frac{m_{\bar{N}}}{10^{9}{\rm GeV}}\right)^{-1} \,.
\label{eq:Yphi}
\eea

Now equating Eq.~(\ref{eq:YDM}) with Eq.~(\ref{eq:Yphi}) and setting $h=0.68$ yield
\be
y\simeq10^{-2}\times\left(\frac{m_{\bar{N}}}{10^{9}{\rm GeV}}\right)^{1/4}\left(\frac{m_{A'}}{1{\rm keV}}\right)^{-1/4}\,.
\label{eq:y2}
\ee
For a given ($m_{\bar{N}}$,$m_{A'}$), as far as $y$ satisfies Eq.~(\ref{eq:y2}), the longitudinal mode of $A_{\mu}^{'}$ ($\theta$) can explain the current DM relic density and $\phi$ is non-thermally produced from $\overline{N}$-$\overline{N}$ scattering by satisfying $\Gamma(\overline{N}+\overline{N}\rightarrow\phi+\phi)<\mathcal{H}$. Note that with this $y$, the condition (3a) in Eq.~(\ref{eq:3a2}) can be rewritten as
\be
{\rm (3a')}:\quad2.5\times10^{-5}\times\left(\frac{m_{\bar{N}}}{10^{9}{\rm GeV}}\right)^{1/2}\left(\frac{m_{A'}}{1{\rm keV}}\right)^{-1/2}>\!\!>\lambda\,.
\label{eq:3asec3}
\ee
Combining the other expression of $y$ in terms of $m_{\phi}$, $m_{\bar{N}}$ and $\lambda$ coming from $m_{\bar{N}}=yV_{\rm B-L}/\sqrt{2}$ with Eq.~(\ref{eq:y2}) gives
\be
10^{-2}\times\left(\frac{m_{\bar{N}}}{10^{9}{\rm GeV}}\right)^{-3/4}\left(\frac{m_{A'}}{1{\rm keV}}\right)^{-1/4}=2\sqrt{\lambda}\left(\frac{m_{\phi}}{10^{9}{\rm GeV}}\right)^{-1}\,,
\label{eq:YADM}
\ee
which is shown in Fig.~\ref{fig:2} as the green dashed lines. For a given ($m_{\bar{N}}$,$m_{A'}$), a set ($\lambda,m_{\phi}$) satisfying Eq.~(\ref{eq:YADM}) makes $A_{\mu}^{'}$ explain the current DM abundance alone (Below the green dashed line, the universe is overclosed by $A_{\mu}^{'}$ relic density).

Now given the conditions (1a), (2a), (3a$'$), (1b) and $\tau_{A'}\gtrsim13.8{\rm Gyr}$ (the age of the universe) where $\tau_{A'}$ is the life time of $A_{\mu}^{'}$, we search for a parameter space in the plane of $(\lambda,m_{\phi})$ satisfying the conditions. In Fig.~\ref{fig:2}, we show the result for the selective sets of $(m_{A'}, T_{\rm RH}, m_{\bar{N}})$. In the figure, the yellow shaded region satisfies the conditions (1a), (2a), (3a$'$) and (1b) while the region above the blue dashed line (the life time condition, i.e. $\tau_{A'}\gtrsim13.8{\rm Gyr}$) makes $\tau_{A'}$ longer than the age of universe.\footnote{Note that Fig.~\ref{fig:2} lacks the red dotted line which appears in Fig.~\ref{fig:3} because $m_{A'}$ in Eq.~(\ref{eq:mmap}) is independent of ($\lambda,m_{\phi}$). Namely, the Lyman-$\alpha$ constraint on $m_{A'}$ is not affected by ($\lambda,m_{\phi}$).} Moreover, the region above the green dashed line makes $A_{\mu}^{'}$ a candidate explaining a fraction of the current DM population. Thus the parameter space for having DPDM from the dark thermal bath is defined to be the part of yellow region lying above the blue and green dashed lines.

In searching for the desired parameter space, we observe that DPDM from the dark thermal bath requires at least $T_{\rm RH}\gtrsim10^{11}{\rm GeV}$. Otherwise, DPDM satisfying the conditions (1a), (2a), (3a$'$) and (1b) cannot live long enough to survive until today. Moreover, $m_{\bar{N}}$ is found to have to be close to $T_{\rm RH}$ due to the conditions (1a) and (1b). The larger gap between $m_{\bar{N}}$ and $T_{\rm RH}$ makes yellow region vertically narrower.\footnote{Note that $\overline{N}$ is relativistic particle as far as $m_{\bar{N}}\lesssim T_{\rm RH}$ since the thermally averaged momentum of $\overline{N}$ is $\sim3.15T_{\rm RH}$.} Eventually we found that for $m_{\bar{N}}=\mathcal{O}(10^{10}){\rm GeV}$, $T_{\rm RH}=\mathcal{O}(10^{11})-\mathcal{O}(10^{12}){\rm GeV}$, $A_{\mu}^{'}$ with $20\lesssim m_{A'}\lesssim30{\rm keV}$ can be the DPDM candidate with $(\lambda,m_{\phi})=(\mathcal{O}(10^{-5}),\mathcal{O}(10^{10}){\rm GeV})$.   

At $T_{\rm DS}\simeq m_{\phi}$, $\phi$ starts to be Boltzmann suppressed, making $A_{\mu}^{'}$ free particle. Defining $a_{\rm FS}$ to be the scale factor at which $A_{\mu}^{'}$ starts its free-propagation, we obtain $a_{\rm FS}\simeq(10^{-13}{\rm GeV})/(m_{\phi}\xi^{-1})$.\footnote{We used $a_{\rm FS}T_{\rm SM}(a_{\rm FS})=a_{\rm EW}T_{\rm SM}(a_{\rm EW})=10^{-13}{\rm GeV}$ with $a_{\rm EW}\simeq10^{-15}$ and $T_{\rm SM}(a_{\rm EW})\simeq100{\rm GeV}$.} In addition, as the phase space of $A_{\mu}^{'}$ is subject to the thermal distribution, its momentum at $a\!=\!a_{\rm FS}$ is $<\!\!p_{A'}(a_{\rm FS})\!\!>\sim2.7\times m_{\phi}$. Since $a_{\rm FS}\times<\!\!p_{A'}(a_{\rm FS})\!\!>$ (and thus $<\!\!p_{A'}(a_{\rm BBN})\!\!>$) does not depend on $m_{\phi}$, we see that both the free-streaming length ($\lambda_{\rm FS}$) in Eq.~(\ref{eq:FSL}) and $\Delta N_{\rm eff}^{\rm BBN}$ in Eq.~(\ref{eq:Neff}) are insensitive to ($\lambda,m_{\phi}$), but sensitive solely to $m_{A'}$.\footnote{$\lambda_{\rm FS}$ still depends on $m_{\phi}$ via the lower limit of the integral in Eq.~(\ref{eq:FSL}). Nevertheless this dependence turns out to be practically irrelevant since $\lambda_{\rm FS}$ is mostly determined by the later time contribution.} Using Eq.~(\ref{eq:FSL}), we figure out that $m_{A'}\simeq20-30$keV gives rise to $\lambda_{\rm FS}=\mathcal{O}(10^{-2})$Mpc. Therefore, the scenario discussed in this subsection is expected to produce the warm DPDM.\footnote{For DM classification based on $\lambda_{\rm FS}$, we refer to Ref.~\cite{Merle:2015oja}. DM travelling $\lambda_{\rm FS}\simeq\mathcal{O}(10^{-2}){\rm Mpc}$ ($\mathcal{O}(10^{-3}){\rm Mpc}$) is defined to be WDM (CDM).} For $m_{A'}$ of interest here, $A_{\mu}^{'}$ behaves as the radiation at BBN era by satisfying $T_{\rm DS}(a_{\rm BBN})>m_{A'}$. So based on Appendix.~\ref{appC}, we check $\Delta N_{eff}^{\rm BBN}$ contributed by DPDM at the BBN era. We find that DPDM of $m_{A'}=\mathcal{O}(10){\rm keV}$ results in $\Delta N_{\rm eff}^{\rm BBN}=\mathcal{O}(10^{-2})$ which is consistent with $\Delta N_{\rm eff}^{\rm BBN}\!\leq\!0.364$~\cite{Cyburt:2015mya}(95\% C.L.).

\section{Without dark thermal bath}
\label{sec4}
In this section, we study the case where the dark sector history is featured by the absence of the dark thermal bath. We envision the situation where $\phi$ free-streams after its production from $\overline{N}$-$\overline{N}$ $t$-channel scattering until it becomes non-relativistic. Afterwards, the non-relativistic free $\phi$ decays to a pair of $A_{\mu}^{'}$s when the time becomes comparable to the life time of $\phi$. Since then, $A_{\mu}^{'}$ free-streams to become DPDM today. 

For this scenario to work, the times corresponding to three events should be compared: the time when (a) $\phi$ becomes non-relativistic, say $a=a_{\rm NR}$, (b) the time when the decay rate of the non-relativistic $\phi$ becomes comparable to the Hubble expansion rate, say $a=a_{\rm FS}$, (c) the time when the dark thermal bath of relativistic $\phi$ and $A_{\mu}^{'}$ can form. For (c), there can be two relevant interactions: $\phi+\phi\leftrightarrow\phi+\phi$ thanks to the quartic term in Eq.~(\ref{eq:LPhi}) and $\phi+\phi\leftrightarrow\theta+\theta$ due to the kinetic term of $\Phi$. For the later process, the interaction rate is approximately $\Gamma\simeq T_{\rm DS}^{5}/V_{\rm B-L}^{4}$ if there is the dark thermal bath with the temperature $T_{\rm DS}$, and it suffices to require $\Gamma<\mathcal{H}$ at $T_{\rm SM}\simeq m_{\bar{N}}$ because most of $\phi$ is produced then and $\Gamma<\mathcal{H}$ lasts for lower $T_{\rm SM}$. We confirmed that the inequality is indeed satisfied for the consistent parameter space which we will show later. Thus, we focus on the self-interaction due to the quartic term. We call the scale factor at which $\Gamma(\phi+\phi\leftrightarrow\phi+\phi)>\mathcal{H}$ holds $a_{\lambda}$. Now when the following hierarchical relation is ensured (condition 1c),
\be
({\rm 1c}):\quad T_{\rm SM}(a_{\rm NR})>T_{\rm SM}(a_{\rm FS})>T_{\rm SM}(a_{\lambda})\,,
\label{eq:1c}
\ee
the dark sector would evolve as we imagine. If we define $a_{\lambda}^{'}$ to be the time when $\lambda^{2}T_{\rm SM}\simeq \mathcal{H}(a_{\lambda}^{'})$ holds, it is clear that $T_{\rm SM}(a_{\lambda}^{'})>T_{\rm SM}(a_{\lambda})$ should be the case since the SM sector should be hotter than the dark sector in any case. From $\lambda^{2}T_{\rm SM}\simeq \mathcal{H}(a_{\lambda}^{'})$, we can infer
\be
T_{\rm SM}(a_{\lambda}^{'})\simeq\lambda^{2} M_{P}\,.
\label{eq:TaL}
\ee

\begin{figure*}[htp]
  \centering
  \hspace*{-5mm}
  \subfigure{\includegraphics[scale=0.395]{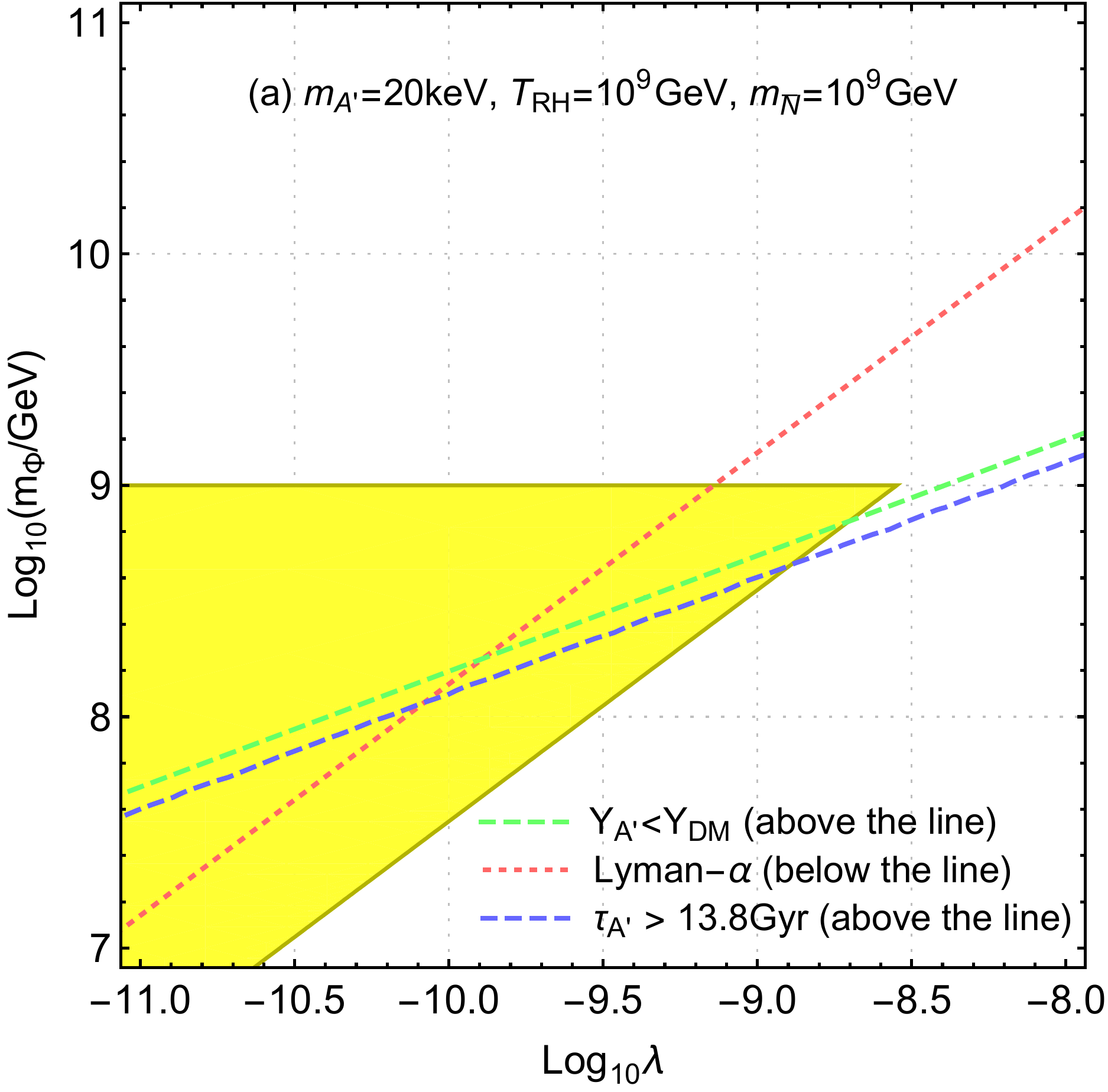}}\qquad
  \subfigure{\includegraphics[scale=0.395]{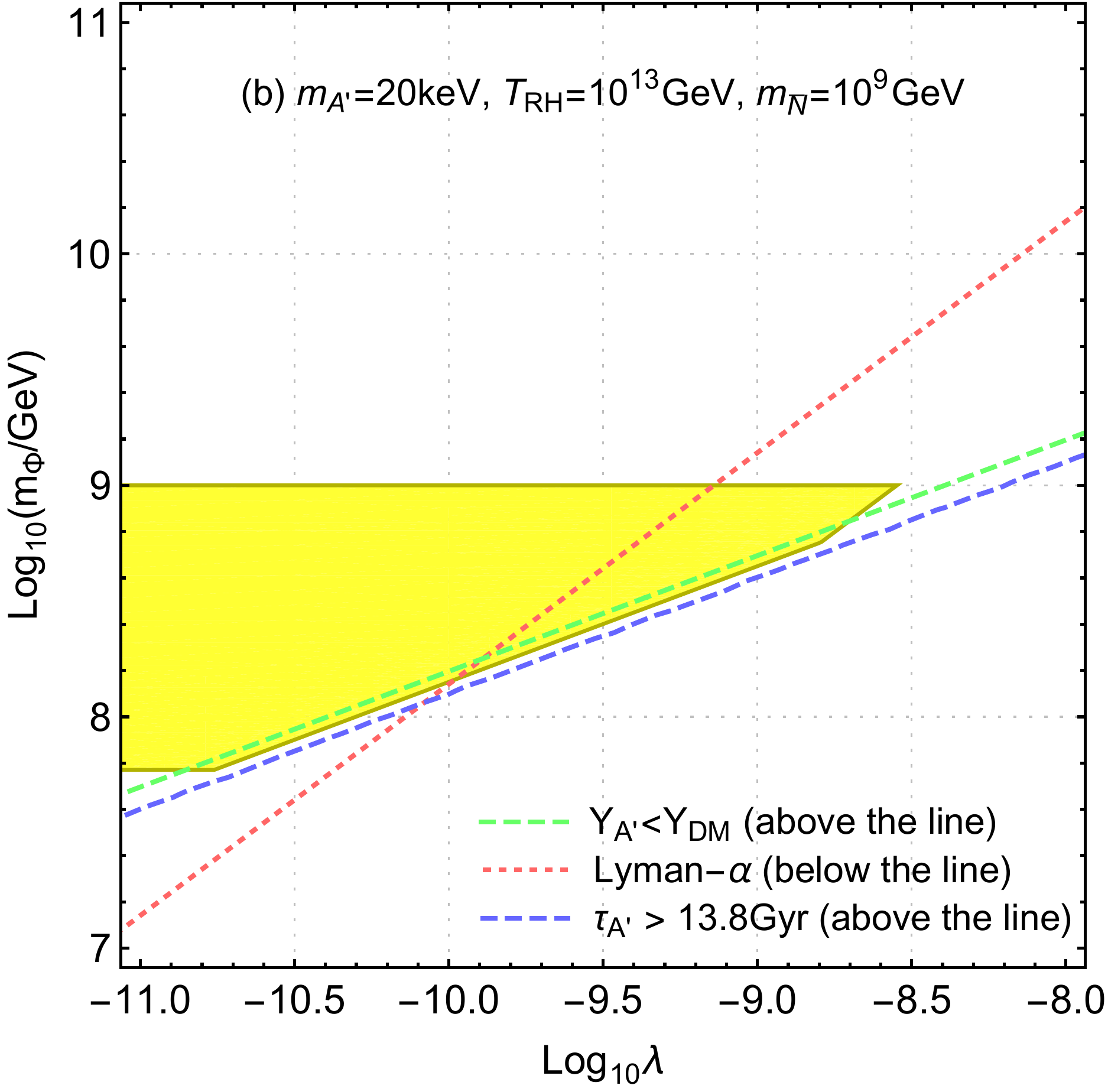}}
  \subfigure{\includegraphics[scale=0.395]{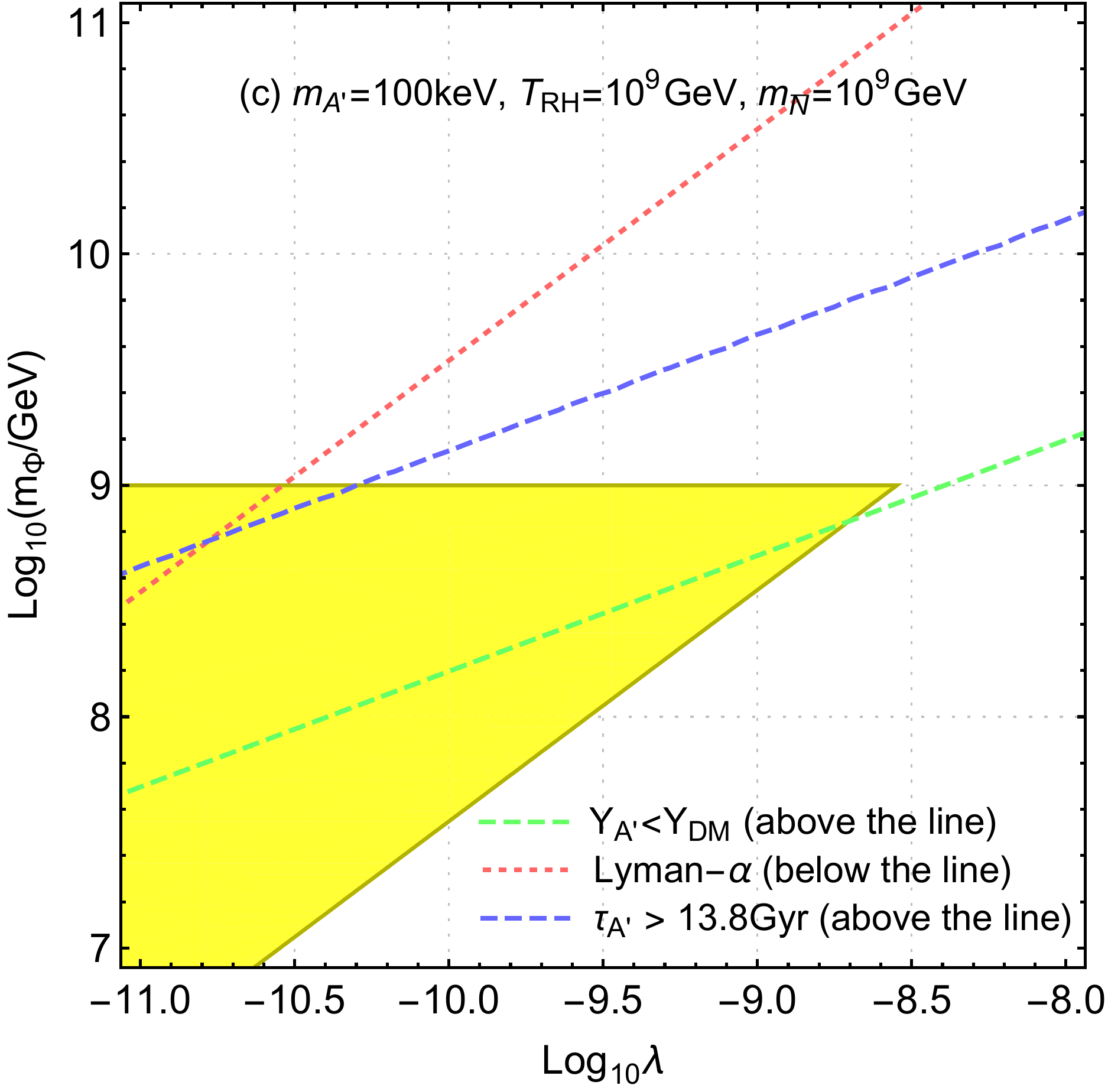}}\qquad
  \subfigure{\includegraphics[scale=0.395]{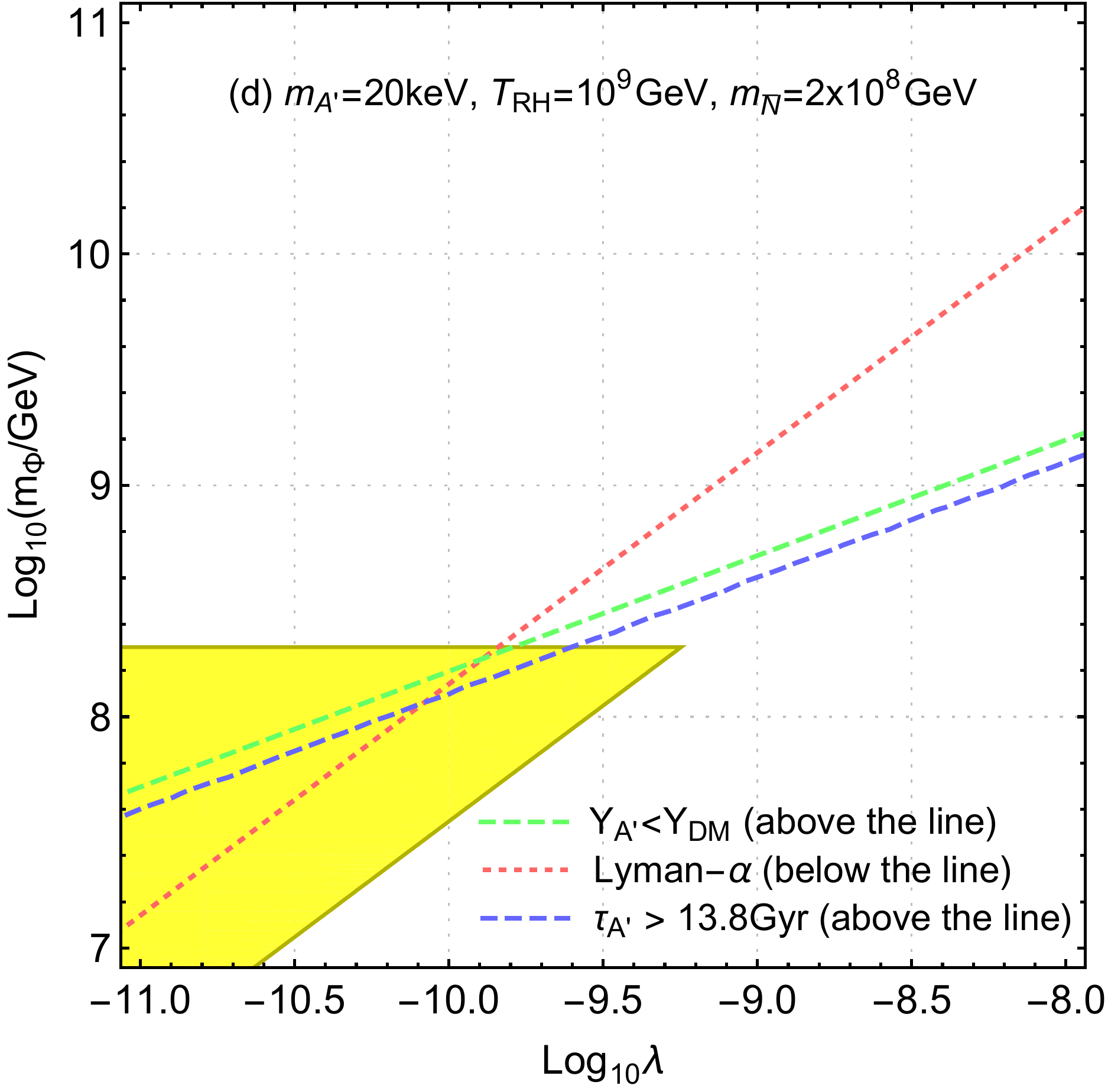}}
  \caption{The constraints on the parameter space of $(\lambda,m_{\phi})$ which is consistent with the conditions (1a), (2a), (3a$''$) and (1c) presented in Eq.~(\ref{eq:VT}), Eq.~(\ref{eq:brokenU(1)}), Eq.~(\ref{eq:3asec4}), Eq.~(\ref{eq:1c}) (yellow shaded region), $\tau_{A'}\gtrsim13.8{\rm Gyr}$ (the region above the blue dashed line) and Lyman-$\alpha$ forest constraint on $m_{A'}$ (the region below the red dotted line). Along the green dashed line, $A_{\mu}^{'}$ alone can explain the current DM abundance. The values shown in (a) being fiducial, each plot of (b), (c) and (d) results from variation of one of ($m_{A'},T_{\rm RH}, m_{\bar{N}}$).}
  \vspace*{-1.5mm}
\label{fig:3}
\end{figure*}

For ensuring Eq.~(\ref{eq:1c}), we demand $T_{\rm SM}(a_{\rm FS})>T_{\rm SM}(a_{\lambda}^{'})$.\footnote{We set this alternative inequality because there is no dark thermal bath considered in this subsection and thus we cannot define $T_{\rm DS}(a_{\lambda})$ and $\xi$.}

Next, for obtaining $T_{\rm SM}(a_{\rm FS})$, we compare the decay rate of $\phi$ to a pair of $A_{\mu}^{'}$s to the Hubble expansion rate,
\bea
&&\Gamma(\phi\rightarrow A^{'}_{\mu}\!+\!A^{'}_{\mu})\simeq\frac{Q_{\rm \Phi}^{4}g_{\rm B-L}^{4}V_{\rm B-L}^{2}m_{\phi}^{3}}{32\pi m_{A'}^{4}}\cr\cr
&&\,\,\quad\qquad\qquad\qquad\simeq\frac{T^{2}_{\rm SM}(a_{\rm FS})}{M_{P}}\simeq \mathcal{H}(a_{\rm FS})\cr\cr&&\Longleftrightarrow\,\,T_{\rm SM}(a_{\rm FS})\simeq2.2\times10^{8}\times\sqrt{\lambda}\times\left(\frac{m_{\phi}}{1{\rm GeV}}\right)^{1/2}{\rm GeV}\,. \nonumber \\
\label{eq:phidecay}
\eea
Using Eq.~(\ref{eq:phidecay}) and $a_{\rm FS}T_{\rm SM}(a_{\rm FS})\!\!\simeq\!\! a_{\rm RH}T_{\rm RH}\!\!\simeq\!\!10^{-13}{\rm GeV}$, we obtain 
\be
a_{\rm FS}\simeq4.6\times10^{-22}\times\lambda^{-1/2}\times\left(\frac{m_{\phi}}{1{\rm GeV}}\right)^{-1/2}\,.
\label{eq:aFS}
\ee

Regarding $T_{\rm SM}(a_{\rm NR})$, we first notice that $\phi$-production is most active at $T_{\rm SM}\!\simeq\! m_{\bar{N}}$ as discussed in Sec.~\ref{sec3}. Let's say this $\phi$-production time $a_{p}$. As $\phi$s are produced from the scattering among $\overline{N}$s that are in the thermal equilibrium, the momentum space distribution of $\phi$ at $a\!=\!a_{p}$ may follow the thermal distribution although $\phi$ is non-thermally produced. Thus $\phi$'s average momentum is expected to be $<\!\!p_{\phi}(a_{p})\!\!>\!\simeq\!2.7\times m_{\bar{N}}$. From $a_{\rm NR}T_{\rm SM}(a_{\rm NR})\!\simeq\! a_{p}T_{\rm SM}(a_{p})\!\simeq\! a_{\rm EW}T_{\rm SM}(a_{\rm EW})\!\simeq\!10^{-13}{\rm GeV}$, we see $T_{\rm SM}(a_{\rm NR})\!\simeq\!10^{-13}{\rm GeV}/a_{\rm NR}$. On the other hand, since $p_{\phi}\!\sim\!1/a$, we have $a_{p}<\!\!p_{\phi}(a_{p})\!\!>\simeq a_{\rm NR}<\!\!p_{\phi}(a_{\rm NR})\!\!>$. Eventually using this information, we obtain
\bea
T_{\rm SM}(a_{\rm NR})&\simeq&\frac{10^{-13}{\rm GeV}}{a_{\rm NR}}\simeq10^{-13}{\rm GeV}\frac{<\!\!p_{\phi}(a_{\rm NR})\!\!>}{a_{p}<\!\!p_{\phi}(a_{p})\!\!>}\cr\cr
&\simeq&\left(\frac{10^{-13}{\rm GeV}}{a_{p}}\right)\left(\frac{m_{\phi}}{2.7m_{\bar{N}}}\right)\cr\cr
&\simeq&\frac{m_{\phi}}{2.7}\,,
\label{eq:TNR}
\eea
where $<\!\!p_{\phi}(a_{\rm NR})\!\!>\simeq m_{\phi}$ and $T_{\rm SM}(a_{p})\simeq m_{\bar{N}}$ are used. By the use of Eq.~(\ref{eq:TaL}), Eq.~(\ref{eq:phidecay}) and Eq.~(\ref{eq:TNR}), we can obtain constraints on $(\lambda,m_{\phi})$ plane which produces DPDM from the decay of the free-streaming $\phi$ based on Eq.~(\ref{eq:1c}).

Since a single $\phi$ decays to two $A_{\mu}^{'}$s, $r\!\equiv\! n_{A'}/n_{\phi}\!=\!2$ applies for this section. Using this number density ratio, the DPDM comoving number density can be written as
\bea
Y_{\rm DM}&\equiv&\frac{n_{A'}}{s_{\rm SM}}=\frac{2n_{\phi}}{s_{\rm SM}}\sim\left.\frac{2n_{\bar{N}}\Gamma(2\overline{N}\rightarrow2\phi)/\mathcal{H}}{s_{\rm SM}}\right\vert_{T_{\rm SM}=m_{\bar{N}}}\cr\cr
&\simeq&3\times10^{5}\times y^{4}\times\left(\frac{m_{\bar{N}}}{10^{9}{\rm GeV}}\right)^{-1} \,.
\label{eq:Yphi2}
\eea
When compared to Eq.~(\ref{eq:YDM}) with $h=0.68$, Eq.~(\ref{eq:Yphi2}) produces
\be
y\simeq9\times10^{-3}\times\left(\frac{m_{\bar{N}}}{10^{9}{\rm GeV}}\right)^{1/4}\left(\frac{m_{A'}}{1{\rm keV}}\right)^{-1/4}\,.
\label{eq:y22}
\ee
Now by Eq.~(\ref{eq:y22}), the condition (1c) in Eq.~(\ref{eq:3a2}) can be rewritten as
\be
{\rm (3a'')}:\quad2\times10^{-5}\times\left(\frac{m_{\bar{N}}}{10^{9}{\rm GeV}}\right)^{1/2}\left(\frac{m_{A'}}{1{\rm keV}}\right)^{-1/2}>\!\!>\lambda\,.
\label{eq:3asec4}
\ee
As with Eq.~(\ref{eq:YADM}), when combined with the expression of $y$ in terms of $m_{\phi}$, $m_{\bar{N}}$ and $\lambda$ coming from $m_{\bar{N}}=yV_{\rm B-L}/\sqrt{2}$, Eq.~(\ref{eq:y22}) yields
\be
9\times10^{-3}\times\left(\frac{m_{\bar{N}}}{10^{9}{\rm GeV}}\right)^{-3/4}\left(\frac{m_{A'}}{1{\rm keV}}\right)^{-1/4}=2\sqrt{\lambda}\left(\frac{m_{\phi}}{10^{9}{\rm GeV}}\right)^{-1}\,,
\label{eq:YADM2}
\ee
which is shown in Fig.~\ref{fig:3} as the green dashed lines. For a given ($m_{\bar{N}}$,$m_{A'}$), a set of ($\lambda,m_{\phi}$) satisfying Eq.~(\ref{eq:YADM2}) makes $A_{\mu}^{'}$ explain the current DM abundance alone (Below the green dashed line, the universe is overclosed by $A_{\mu}^{'}$ relic density).

Because $A_{\mu}^{'}$ has never the chance to reside in the dark thermal bath, its phase space distribution does not follow the usual thermal distribution. Rather, it is subject to $f(q,t)=(\alpha/q){\rm exp}(-q^{2})$ with $q\equiv p/T$~\cite{Kaplinghat:2005sy,Strigari:2006jf,Aoyama:2011ba,Kamada:2013sh,Merle:2013wta} and $\alpha$ a normalization constant. With this $f(q,t)$ of $A_{\mu}^{'}$, we find $\tilde{\sigma}\simeq1$ based on Eq.~(\ref{eq:warmness}). Also from $<\!\!p_{A'}(a_{\rm FS})\!\!>\simeq m_{\phi}/2$ and Eq.~(\ref{eq:aFS}), we find the DPDM temperature today to be 
\bea
T_{A',0}&\simeq&a_{\rm FS}<\!\!p_{A'}(a_{\rm FS})\!\!>\cr\cr
&\simeq&2.3\times10^{-22}\times\lambda^{-1/2}\times\left(\frac{m_{\phi}}{1{\rm GeV}}\right)^{1/2}{\rm GeV}\,.
\label{eq:TAprime}
\eea
Invoking $\tilde{\sigma}\simeq1$, Eq.~({\ref{eq:TAprime}}), Eq.~(\ref{eq:mmap}) and $m_{\rm wdm}>5.3{\rm keV}$ (constraint on the thermal WDM mass)~\cite{Irsic:2017ixq}, we find (condition 2c)
\be
({\rm 2c}):\quad m_{A'}>1.7\times10^{-8}\times\lambda^{-1/2}\times\left(\frac{m_{\phi}}{1{\rm GeV}}\right)^{1/2}{\rm keV}\,,
\label{eq:2c}
\ee
which makes the presence of $A_{\mu}^{'}$ consistent with the Lyman-$\alpha$ forest observation.

In accordance with the conditions (1a), (2a), (3a$''$), (1c), (2c), $Y_{A'}\lesssim Y_{\rm DM}$ and $\tau_{A'}\gtrsim13.8{\rm Gyr}$, we probe a parameter space in the plane of $(\lambda,m_{\phi})$ satisfying the conditions. We show the result in Fig.~\ref{fig:3} for the selective sets of $(m_{A'}, T_{\rm RH}, m_{\bar{N}})$. In the figure, the yellow shaded region satisfies the conditions (1a), (2a), (3a$''$) and (1c). The region above the blue dashed line  makes $\tau_{A'}$ longer than the age of universe. In addition, the region above the green dashed line makes $A_{\mu}^{'}$ explain a fraction of the current DM population. The region below the dotted red line ensures that DPDM from the decay of the non-relativistic free $\phi$ is consistent with the observation of Lyman-$\alpha$ forest data. The viable parameter space is defined to be the part of yellow region lying above the blue and green dashed lines, and below the red dotted line.

In searching for the viable parameter space, we observe that DPDM from the decay of the non-relativistic free $\phi$ requires at least $T_{\rm RH}\gtrsim3\times10^{8}{\rm GeV}$. For this $T_{\rm RH}\simeq3\times10^{8}{\rm GeV}$, even $m_{A'}$ as low as $15$keV can be marginally viable for the narrow parameter space. When $T_{\rm RH}$ increases with other parameters fixed, due to the condition (2a), the possible lower bound of $m_{\phi}$ increases so that the area of a viable region decreases as shown the panel (b) in Fig.~\ref{fig:3}. Put it another way, as separation between $T_{\rm RH}$ and $m_{\bar{N}}$ increases, the area of a viable parameter space tends to decrease. However, importantly if increase in $T_{\rm RH}$ is accompanied by increase in $m_{\bar{N}}$, then the yellow shaded region moves upward making the area of a viable parameter space increase. Increasing $m_{A'}$ moves both the blue dashed and red dotted line upward and to the left, implying that there exists a upper bound for $m_{A'}$ for a fixed $T_{\rm RH}$. We find that the allowed range of $m_{A'}$ is $11\sim150$keV for $m_{\bar{N}}\simeq T_{\rm RH}\simeq10^{9}{\rm GeV}$ while $m_{A'}$ can be as large as 1MeV when $m_{\bar{N}}\simeq T_{\rm RH}\simeq10^{11}{\rm GeV}$. We observe that in order for $A_{\mu}^{'}$ to explain the current DM abundance alone, for a given set of $(T_{\rm RH},m_{\bar{N}})$, the smaller $m_{A'}$ is preferred.

For the viable parameter spaces we can obtain by varying $(m_{A'},T_{\rm RH},m_{\bar{N}})$, we checked $\lambda_{\rm FS}$ and $\Delta N_{\rm eff}^{\rm BBN}$, using $a_{\rm FS}$ in Eq.~(\ref{eq:aFS}) and $<\!\!p_{A'}(a_{\rm FS})\!\!>\simeq m_{\phi}/2$. The parameter spaces satisfy the constraint $\Delta N_{\rm eff}^{\rm BBN}\!\leq\!0.364$~\cite{Cyburt:2015mya}(95\% C.L.) readily by yielding $\Delta N_{\rm eff}^{\rm BBN}=\mathcal{O}(10^{-2})$. As for the $\lambda_{\rm FS}$, differing from the scenarios discussed in Sec.~\ref{sec3}, we find that DPDM can be characterized by both the warm and cold nature based on the scenario studied in Sec.~{\ref{sec4}}. The diverse range of $\lambda_{\rm FS}\sim\mathcal{O}(10^{-3})-\mathcal{O}(10^{-2})$Mpc is produced and the larger $m_{\phi}$ and the smaller $\lambda$ tend to make $\lambda_{\rm FS}$ longer. This tendency can be understood by observing Eq.~(\ref{eq:aFS}) and Eq.~(\ref{eq:FSL}): for a fixed $m_{\phi}$ and hence $<\!\!p_{\rm A'}(a_{\rm FS})\!\!>$, the smaller $\lambda$ delays the onset of the free-streaming of DPDM so as to induce increase in the infinitesimal contribution d$\lambda_{\rm FS}$ at the same time of $a>a_{\rm FS}$. This results in the overall increase in $\lambda_{\rm FS}$ since the later time makes more contribution to $\lambda_{\rm FS}$ than the earlier time. We notice that the absence of the dark thermal bath in this section requires $\lambda$ to be much smaller than considered in Sec.~{\ref{sec3}} and thus becomes the main reason to make WDM possibility in this section viable. On the other hand, increasing $m_{\phi}$ leads on to increase in $<\!\!p_{A'}(a_{\rm FS})\!\!>\simeq m_{\phi}/2$ so that $\lambda_{\rm FS}$ becomes longer.

\section{Conclusion}
In this paper, we probed the possibility in which the gauge boson ($A_{\mu}^{'}$) of the gauged $U(1)_{\rm B-L}$ symmetry is identified with the current DM population. As the keV-scale DM candidate, it is required to introduce the dark sector system cooler than and isolated from the SM sector thermal bath. For producing such a dark sector system, we invoked the Yukawa coupling between $\Phi$ and $\overline{N}$ of which existence is unquestionable in the light of the seesaw mechanism and the leptogenesis. The scalars are produced non-thermally from scattering among $\overline{N}$s in the SM thermal bath. The seesaw mechanism and the leptogenesis can be easily realized by the decay of two other right-handed neutrinos, i.e. $\overline{N}_{1}$ and $\overline{N}_{2}$. The leptogenesis could be either thermal or non-thermal one depending on masses of $\overline{N}_{1}$ and $\overline{N}_{2}$. Depending on the Yukawa interaction strength, we find that there could be two distinguished ways of evolution for the dark sector system: evolution with (Sec.~\ref{sec3}) and without the dark thermal bath (Sec.~\ref{sec4}).

For the case with the dark thermal bath, we find that the thermal bath is bound to include both the scalar and $A_{\mu}^{'}$. After the heavier scalar is integrated-out, $A_{\mu}^{'}$ starts free-streaming until today, becoming the DM candidate in the model. For the case without the dark thermal bath, the scalar continues to free-stream after its production until it becomes non-relativistic particle. Then, the expansion rate of the universe becoming slow enough to be comparable to its decay rate, the scalar starts to disappear by decaying to $A_{\mu}^{'}$s. Thereafter, $A_{\mu}^{'}$ free-streams until today to become DM candidate today. For both situations, the small enough gauge coupling constant $g_{\rm B-L}$ makes $A_{\mu}^{'}$'s free-streaming undisturbed to date since the decay of the scalar.

For both cases, the DPDM life time constraint (blue dashed lines shown in Fig.~\ref{fig:2} and Fig.~\ref{fig:3}) apply equally, requiring a larger $m_{\phi}$ for a stronger strength of $\lambda$. We notice that the required $\lambda$ for the case with the dark thermal bath is stronger than the case without the dark thermal bath, implying that a consistent $m_{\phi}$ should be larger for the case with the dark thermal bath. Since $m_{\phi}$ should be at least smaller than a reheating temperature ($T_{\rm RH}$) for non-thermal production of $\phi$, we see that the case with the dark thermal bath is consistent with the high $T_{\rm RH}$ (as high as $T_{\rm RH}\simeq10^{11}{\rm GeV}$) whereas the case without the dark thermal bath can produce DPDM even for $T_{\rm RH}$ as low as $\mathcal{O}(10^{8})$GeV.

As we pointed out in Sec.~\ref{sec:model}, the life time constraint results in $g_{\rm B-L}<\mathcal{O}(10^{-18})$ for the keV-scale $m_{A'}$. This makes it very challenging to detect $A_{\mu}^{'}$ via the electron recoil in experiments like Xenon1T or PandaX. However, if there exists a non-vanishing kinetic mixing between $U(1)_{\rm B-L}$ gauge boson and the SM photon, we may have signals of the electron recoil for $10-100$keV energy regime in those experiments (see, e.g. \cite{Choi:2020udy}). As regards the small gauge coupling, we further notice that the model is featured by $V_{\rm B-L}\geq\mathcal{O}(10^{12}){\rm GeV}$ for $m_{A'}$ of our interest. This is interesting to be very consistent with the standard seesaw mechanism.

Differing from the previous discussions about a DM candidate in the minimal $B-L$ model, to our best knowledge, the first proposal of identifying $U(1)_{\rm B-L}$ gauge boson with the DM candidate was made in this paper. As shown thus far, aside from the Yukawa coupling $y$ and the scalar sector parameters, quantities in the model including $T_{\rm RH}$ and the right-handed neutrino mass $m_{\bar{N}}$ are involved with the consistent parameter space for having $U(1)_{\rm B-L}$ gauge boson as the DM candidate. Although we have not made discussion about an experimental way to search for DPDM discussed in this paper, if there appears a novel way to probe DPDM in $B-L$ model, we expect it to provide us with an indirect way of probing $T_{\rm RH}$ and $m_{\bar{N}}$ because of the correlation between consistent parameter spaces for DPDM and these quantities.

We emphasize that the main purpose of this paper is to show the presence of consistent parameter spaces with $A_{\mu}^{'}$ being the DM. We have taken some special assumptions for couplings and role of three right handed neutrinos in order to make our analysis simpler. However, we can perform similar analysis by taking more general couplings of the right-handed neutrinos so as to treat all contributions for $\phi$-production shown in Fig.~\ref{fig} on an equal footing. Moreover, we may have another mechanism for $\phi$-production other than the use of the operator $\sim\Phi\overline{N}\overline{N}$. We leave these extra general analysis for future work.


\begin{acknowledgments}
N. Y. is supported by JSPS KAKENHI Grant Numbers JP16H06492.
T. T. Y. is supported in part by the China Grant for Talent Scientific Start-Up Project and the JSPS Grant-in-Aid for Scientific Research No. 16H02176, No. 17H02878, and No. 19H05810 and by World Premier International Research Center Initiative (WPI Initiative), MEXT, Japan. 

\end{acknowledgments}


\appendix
\section{DM relic abundance}
\label{appA}
The comoving number density of DM ($Y_{\rm DM}\equiv n_{\rm DM}/s$) is related to the DM relic abundance today via
\be
\Omega_{\rm DM,0}=\frac{\rho_{\rm DM,0}}{\rho_{\rm cr,0}}=\frac{s_{0}}{\rho_{\rm cr,0}}\times m_{\rm DM}\times Y_{\rm DM}\simeq0.27\,,
\label{eq:OmegaDM}
\ee
where the entropy density and the critical density today are given by $s_{0}=2.21\times10^{-11}{\rm eV}^{3}$ and $\rho_{\rm cr,0}=8.03\times10^{-11}\times h^{2}\,{\rm eV}^{4}$ respectively. Here $h$ is defined through $\mathcal{H}_{0}=100h{\rm km/sec/Mpc}$. Rewriting $Y_{\rm DM}$ in terms of $m_{\rm DM}$ based on Eq.~(\ref{eq:OmegaDM}), we obtain
\begin{equation}
    Y_{\rm DM}=9.8\times10^{-4}\times\left(\frac{m_{\rm DM}}{1{\rm keV}}\right)^{-1}\times h^{-2}\,.
\label{eq:YDM}
\end{equation}

For the case with the ratio $r=n_{A'}/n_{\Phi}$, the temperature ratio $\xi\equiv T_{\rm DS}/T_{\rm SM}$ at the time when $\phi$-production is most active reads
\be
\xi\simeq0.93\times r^{-1/3}\times\left(\frac{m_{\rm DM}}{1{\rm keV}}\right)^{-1/3}\,.
\label{eq:xi}
\ee

\section{Free-streaming length}
\label{appB}
When a particle becomes free at the time of the scale factor $a_{\rm FS}$ and its average momentum at the time reads $<\!\!p_{\rm DM}(a_{\rm FS})\!\!>$, the distance the particle travels since then is estimated by
\bea
\lambda_{\rm FS}&=&\int_{t_{\rm FS}}^{t_{0}}\frac{<\!\!v_{\rm DM}(t)\!\!>}{a}{\rm d}t\cr\cr
&=&\int_{a_{\rm FS}}^{1}\frac{{\rm d}a}{H_{0}\Omega(a)}\frac{<\!\!p_{\rm DM}(a_{\rm FS})\!\!>a_{\rm FS}}{\sqrt{(<\!\!p_{\rm DM}(a_{\rm FS})\!\!>a_{\rm FS})^{2}+m_{\rm DM}^{2}a^{2}}}\,, \nonumber \\
\label{eq:FSL}
\eea
where $\Omega(a)\equiv\sqrt{\Omega_{\rm rad,0}+a\Omega_{\rm m,0}+a^{4}\Omega_{\Lambda,0}}$ is defined and used. Here $\Omega_{x,0}\equiv\rho_{x,0}/\rho_{\rm cr,0}$ denotes the fraction of the critical density attributed to the current energy density of the species $x$ (radiation, matter and dark energy). For an estimation of $\lambda_{\rm FS}$, we use $\Omega_{\rm rad,0}=9.4\times10^{-5}$, $\Omega_{\rm m,0}=0.32$ and $\Omega_{\Lambda,0}=0.68$.

\section{$\Delta N_{\rm eff}$}
\label{appC}
If a DM candidate is still relativistic at the BBN era ($T_{\rm SM}(a_{\rm BBN})\simeq1{\rm MeV}$), it behaves as the radiation and makes extra contribution to $N_{\rm eff}^{\rm BBN}$ on top of the SM neutrino contribution $N_{\rm eff}^{\rm BBN}=3.046$. This additional contribution is parametrized by 
\be
\Delta N_{\rm eff}^{\rm BBN}\simeq\frac{\rho_{{\rm DM}}(a_{\rm BBN})}{\rho_{\gamma}(a_{\rm BBN})}\times\frac{8}{7}\left(\frac{11}{4}\right)^{4/3}\,,
\label{eq:Neff}
\ee
where $\rho_{\gamma}(a_{\rm BBN})$($\rho_{\rm DM}(a_{\rm BBN})$) is the energy density of the photon (DM) at the BBN time. The energy density of DM can be estimated to be
\bea
\rho_{{\rm DM}}(a_{\rm BBN})&=&\sqrt{m_{{\rm DM}}^{2}+<\!\!p_{\rm DM}(a_{\rm BBN})\!\!>^{2}}\times Y_{\rm DM}\cr\cr
&&\times\frac{2\pi^{2}}{45}g_{s,{\rm SM}}(a_{{\rm BBN}})T_{{\rm SM}}(a_{\rm BBN})^{3}\,,
\label{eq:rhoDM}
\eea
where $g_{\rm s,SM}$ is the effective number of relativistic degrees of freedom for the entropy density and $Y_{\rm DM}$ can be read from Eq.~(\ref{eq:YDM}).

\section{WDM mass constraint mapping}
\label{appD}
By identifying the warmness $\sigma$ of the usual thermal WDM and the WDM candidate in a specific model, one can obtain the lower bound of the WDM candidate's mass in the model consistent with the Lyman-$\alpha$ forest observation~\cite{Kamada:2019kpe}. When the momentum space distribution of the WDM of interest is $f(q)$, the corresponding warmness is defined to be
\be
\sigma\equiv\tilde{\sigma}\frac{T_{DM}}{m_{\rm DM}}\quad{\rm with}\quad\tilde{\sigma}^{2}\equiv\frac{\int dq q^{4}f(q)}{\int dq q^{2}f(q)}\,.
\label{eq:warmness}
\ee
Equating the warmness for the the usual thermal WDM ($\sigma_{\rm wdm}$) and for DPDM in our model ($\sigma_{A'}$), we obtain the mass mapping
\be
m_{A'}=\left(\frac{\tilde{\sigma}_{A'}}{3.6}\right)\left(\frac{T_{A',0}}{T_{\rm wdm,0}}\right) m_{\rm wdm}\,,
\label{eq:mmap}
\ee
where the present temperature of the usual thermal WDM is given by~\cite{Viel:2005qj}
\be
T_{\rm wdm,0}\simeq\left[0.036\left(\frac{94{\rm eV}}{m_{\rm wdm}}\right)\right]^{1/3}T_{\gamma,0}\,,
\ee
where $T_{\gamma,0}\simeq2.35\times10^{-4}{\rm eV}$ is the present photon temperature.

\bibliography{main}

\end{document}